\documentclass[preprint2]{aastex}

\usepackage{natbib}
\usepackage{graphicx}
\usepackage{multirow}
\usepackage{txfonts}

\begin{document}

\shorttitle{X-ray Flux of Jets of Swift J1753-0127 from TCAF}
\shortauthors{A. Jana, S. K. Chakrabarti, \& D. Debnath}

\title{Properties of X-ray Flux of Jets During 2005 Outburst of Swift J1753.5-0127 Using TCAF Solution}
\author{Arghajit Jana\altaffilmark{1}, Sandip K. Chakrabarti\altaffilmark{2,1}, Dipak Debnath\altaffilmark{1}}
\altaffiltext{1}{Indian Center for Space Physics, 43 Chalantika, Garia St. Rd., Kolkata, 700084, India.}
\altaffiltext{2}{S. N. Bose National Centre for Basic Sciences, Salt Lake, Kolkata, 700106, India.}

\email{argha@csp.res.in, chakraba@bose.res.in, dipak@csp.res.in}

%\date{Accepted; Received }

\begin{abstract}

Galactic black hole candidate Swift~J1753.5-0127 was discovered on 2005 June 30 by the Swift Burst Alert Telescope. 
We study the accretion flow properties during its very first outburst through careful analysis of the evolution 
of the spectral and the temporal properties using the two-component advective flow (TCAF) paradigm. RXTE 
proportional counter array spectra in $2.5-25$~keV are fitted with the current version of the TCAF model fits 
file to estimate physical flow parameters, such as two component (Keplerian disk and sub-Keplerian halo) 
accretion rates, properties of the Compton cloud, probable mass of the source, etc. The source 
is found to be in harder (hard and hard-intermediate) spectral states during the entire phase of the outburst with 
very significant jet activity. 
Since in TCAF solution, the model normalization is constant for any particular source, any requirement of significantly different 
normalization to have a better fit on certain days would point to X-ray contribution from components not
taken into account in the current TCAF model fits file. By subtracting the contribution using actual normalization, 
we derive the contribution of X-rays from the jets and outflows. We study its properties, such as its magnitude and spectra.
We find that on some days, up to about 32\% X-ray flux is emitted from the base of the jet itself.

\end{abstract}


\section{Introduction}

Stellar mass black holes candidates (BHCs) exhibiting transient behavior generally reside in binaries. They show occasional outbursts 
of variable duration ranging from few weeks to months. In between two outbursts, these transient BHCs stay in 
long periods of quiescence. During the outbursts, compact objects (here, BHCs) accrete matter from their companions 
via Roche-lobe overflow and/or by wind accretion, which forms a disk-like structure, commonly known as an {\it accretion disk}. 
Electromagnetic radiation from radio to $\gamma$-rays are emitted from the disk, which  makes it observable.
It is believed that an outburst is triggered by a sudden rise in viscosity in the disk, which increased 
the accretion rates in the inner disk causing outbursts (Chakrabarti, 2013). Rapid evolution of spectral and temporal properties are 
observed during an outburst of transient BHCs and these are found to be strongly correlated. In the hardness-intensity 
diagram (HID; Fender et al. 2004; Debnath et al. 2008) or accretion rate ratio intensity diagram (ARRID; Jana et al. 2016), 
observed in different states are found to be correlated with different branches.
Generally four spectral states, namely, the hard (HS), hard-intermediate (HIMS), soft-intermediate (SIMS) and soft (SS) 
states are observed during an outburst. Each state is defined with certain characteristics of spectral and temporal features. 
HS and HIMS are dominated by non-thermal high energetic radiations with observation of monotonical rise/fall of low frequency 
quasi-periodic oscillations (QPOs), whereas SIMS and SS are dominated by thermal radiations with sporadic QPOs (in SIMS) 
or no QPOs (in SS) (for more details, see Nandi et al. 2012; Debnath et al. 2010, 2013 and references therein). 
According to Debnath et al. (2017), outbursts are of two types: type-I or classical type, where 
all spectral states are observed, and type-II or harder type, where SS are absent. The latter type of outbursts 
are termed as `failed' outbursts. For instance, 2005 outburst of Swift~J1753.5-0127 is of type-II.

Black hole (BH) X-ray spectrum consists of both thermal and non-thermal components. The thermal component is basically a
multicolor blackbody that is emitted from the standard Keplerian disk (Shakura \& Sunyaev 1973). The non-thermal 
component is of power-law (PL) type, and it originates from the so-called `hot corona' or 'Compton cloud' (Sunyaev \& Titarchuk 1980).
In the two-component advective flow (TCAF) solution (Chakrabarti \& Titarchuk 1995), this corona is identified with 
the CENtrifugal pressure supported BOundary Layer (CENBOL), which naturally forms behind the centrifugal barrier due 
to pile-up of the free-falling, weakly viscous (less than critical viscosity), sub-Keplerian (low angular momentum) matter. 
Soft photons from the Keplerian disk gain energy by repeated inverse-Compton scattering with the hot electron at the CENBOL 
and emerge as high energetic photons having a power-law distribution in energy.  
Recently, this TCAF solution has been included in HEASARC's spectral analysis software package XSPEC as an additive table 
model to fit BH spectra (Debnath et al. 2014, 2015a). Few transient BHCs have been studied by our group during their 
X-ray outbursts to find a clear picture about the evolution of the physical properties of these sources
during their X-ray outbursts (Mondal et al. 2014, 2016; Debnath et al. 2015a,b, 2017; 
Chatterjee et al. 2016; Jana et al, 2016; Bhattacharjee et al. 2017; Molla et al. 2017). 

Jets and outflows are important features in accretion disk dynamics. According to the TCAF paradigm, 
the jets and outflows are produced primarily from the CENBOL region (Chakrabarti 1999a; Das \& Chakrabarti 1999).
If this region remains hot as in hard and hard-intermediate states, jets could be produced, otherwise not.
Generally, inflow rates increase as the object goes from the hard state to the hard-intermediate state,
higher outflow rates are also observed in the intermediate states. It is also reported in the literature that blobby-jets 
are possible in intermediate states (Chakrabarti, 1999b; 2001; Nandi et al. 2001) due to higher optical depth
at the base of the jet which episodically cools and separates the jets.
In softer states, this region is quenched and the outflow rates are reduced (also see, Garain et al. 2013). 
Collimation of the jets could be accomplished by toroidal flux tubes emerging from generally convective 
disks (Chakrabarti \& D’Silva 1994; D’Silva \& Chakrabarti 1994). 
There are several papers in the literature that invoke diverse mechanisms for the acceleration 
of this matter discussion of which is beyond the scope of the present paper. In the present paper, 
we introduce a new method to estimate X-ray flux, emitted from the base of the jets during the entire 
period of the 2005 outburst of Swift~J1753.5-0127 and compare that with the radio observations. 

Radio jets are common in active galactic nucleus (AGN) sources. It has been observed for several Galactic BHCs, such as, GRS~1758-258 
(Rodriguez et al., 1992), 1E~1740.7-2942 (Mirabel et al., 1992) etc. Compact radio jets have been detected in BHCs, such as, 
GRS~1915+105 (Dhawan et al., 2000), Cyg X-1 (Stirling et al., 2001). The BHCs GRS~1915+105 (Mirabel \& Rodriguez, 1994) 
and GRO~J1655-40 (Tingay et al., 1995, Hjellming \& Rupen, 1995) show superluminal jets. 
Though jets are prominent in radio, they could be observed in other 
energy bands, such as, X-rays and $\gamma$-rays. High energy $\gamma$-ray jets 
have been observed in Cyg X-1 (Laurent et al. 2011, Jourdain et al. 2012) and V~404~Cyg (Loh et al. 2016).
Large scale, decelerating relativistic X-ray emitting jets have been observed in BHC XTE~J1550-564 
(Corbel et al. 2002a, 2002b, Kaaret et al. 2006, Tomsick et al. 2003). In this case, radio blobs were predicted 
to move at relativistic speed, with blobs emitting in X-rays. H~1743-322 also showed a similar X-ray jet (Corbel et al. 2005). 
Kaaret et al. (2006) reported large scale X-ray jet in BHC 4U~1755-33. 
A relation between IR and X-ray jets has been found in BHC GRS~1915+105 (Eikenberry et al. 1998; 
Lasso-Cabrera \& Eikenberry, 2013). X-ray jet of SS~433 even close to the black hole is well known. 
A correlation between X-ray and radio band intensity in compact jets was first found in BHC 
GX~339-4 (Hannikainen et al. 1998). The standard correlation is $F_{R} \propto F_{X}^b$ with $b \sim 0.6-0.7$ 
(Corbel et al. 2003; Gallo et al. 2003). This empirical relation is thought to be universal, although for some BHCs, it is observed 
to have steeper PL with index $\sim 1.4$ (Jonker et al. 2004; Coriat et al. 2011). Some BHCs also have shown 
dual track in correlation plot. Dual correlation indices were observed for BHCs GRO~J1655-40 (Corbel et al. 2004), H~1743-322 
(Coriat et al. 2011), XTE~J1752-522 (Ratti et al. 2012), MAXI~J1659-152 (Jonker et al. 2012).
Until now, the radio and X-ray correlation study was done using quasi-simultaneous data of radio and X-ray fluxes.
Usually, total X-ray flux (disk plus jet)  is used for the correlation.

It is reported that jets are emitted in the entire range of electromagnetic spectra: radio to $\gamma$-ray. 
Thus X-rays emitted from BHCs when jets are present is the net contribution coming from both the jet and the accretion 
disk. Till now, there was no way to separate the contribution of these two components.
In the present paper, for the first time, we make an attempt to separate these two components from the total observed X-rays 
using the unique aspects of spectral studies by the TCAF solution. These are radiation in the accretion 
disk component is contributed by the Keplerian disk (dominating the soft X-ray band) and 
from the `hot Compton cloud' region, i.e., from the CENBOL (dominating the hard X-ray band) and the normalization can be
treated as a constant across the spectral states.

Swift~J1753.5-0127 is discovered on 2005 June 30 by Swift/BAT instrument at RA$=17^h 53^m 28^s.3$, DEC$=-01^\circ 27' 09''.3$ 
(Palmer et al. 2005). BHC Swift~J1753.5-0127 has a short orbital period ($2.85$~hrs according to Neustroev et al. 2014; 
and $3.2\pm 0.2 $~hrs according to Zurita et al. 2007). Neustroev et al. (2014) also estimated the mass of the source as 
$< 5~M_{\odot}$ and the companion mass to be between $0.17-0.25 M_{\odot}$ with the disk inclination angle $>40^\circ$. 
On contrary, Shaw et al. (2016) estimated the mass as $>7.4~M_{\odot}$. The distance of the source is estimated to be 
$4-8$~kpc (Cadolle Bel et al. 2007).
Radio jets are also observed during 2005 outburst of the source (Fender et al. 2005; Soleri et al. 2010).
Several authors have found radio/X-ray correlation for this source. This does not fall on the traditional correlation track, 
rather, it shows power-law index to be steeper $\sim 1-1.4$ (Soleri et al. 2010, Rushton et al. 2016, Kolehmainen et al. 2016).

In Debnath et al. (2017; hereafter Paper-I), a detailed study of the spectral and temporal properties of this object during 
its 2005 outbursts (from  2005 July 2 to 2005 October 19) was made. They used TCAF model fits file to fit the spectra and 
obtained accretion flow properties of the source during the outburst. Based on the variations of the TCAF model fitted 
(spectral) physical flow parameters and observed QPO frequencies, entire 2005 outburst was classified into two harder 
spectral states, such as, HS \& HIMS, and these states were observed in the sequence: 
HS (Ris.) $\rightarrow$ HIMS (Ris.) $\rightarrow$ HIMS (Dec.) $\rightarrow$ HS (Dec.). 
They also estimated the mass of the BHC to be in the range of $4.75-5.90$~M$_{\odot}$ or $5.35^{+0.55}_{-0.60}$~M$_{\odot}$.
According to the TCAF solution, model normalization (N) is a function of intrinsic parameters, such as, distance, mass 
and constant inclination angle of the binary system. So, $N$ is a constant for a particular BHC across, its spectral states
unless there is a precession in the disk to change the projected emission surface area or there are some significant outflow 
or jet activities which so far are not included in the current version (v0.3) of the TCAF model {\it fits} file. As reported 
in Paper-I, there are significant deviation of the constant $N$ in few observations during the outburst. This allows us to 
estimate the amount of jet flux by separating it from the total X-ray luminosity from our spectral study with the current 
version of the TCAF solution by keeping model normalization frozen at the lowest observed value. The spectral property of the 
residual X-ray is also found.

The {\it paper} is organized in the following way. In \S 2, we briefly discuss the relation of jet with spectral states.
In \S 3, we also briefly present a method to estimate the jet flux from the total 
X-ray flux. In \S 4, we present results on our estimated jet flux and its evolution during the entire 2005 outburst of 
Swift~J1753.5-0127. We compare our estimated jet flux with that of the radio fluxes observed during the outburst 
and study correlation between X-ray and radio jet flux components. 
Finally, in \S 5, a brief discussion and concluding remarks are presented.

\section{Disk-Jet Connection with Spectral States}

In general, there are two types of jets: continuous outflows (Compact jets) and discrete ejections (blobby jets:
 Chakrabarti \& Nandi 2000; Chakrabarti et al. 2002). In TCAF, CENBOL acts as a base of the jet (Chakrabarti 1999a). 
Ejection of the matter depends on the shock location ($X_s$), compression ratio ($R$) and inflow rate. 
A schematic diagram of inflow and outflow is shown in the second panel of Fig. 1. Jet move subsonically up to the 
sonic surface ($\sim 2.5 X_s$) and then moves away supersonically, thereby reducing its temperature during expansion 
and emitting in UV, IR to radio (Chakrabarti 1999ab; Chakrabarti \& Manickam 2000, hereafter CM00). 
The subsonic region will upscatter seed photons  from the Keplerian disk and downscatter CENBOL photons contributing to softer X-rays, 
which we define here as the jet X-ray ($F_{ouf}$) flux in this paper. This does not include the X-rays emitted from 
interaction of the jet with ambient medium. If the CENBOL is not hot, i.e., the object is not in the hard 
or hard intermediate states, compact jets are not possible. However, as the shock moves in due to 
larger inflow rates and consequent post-shock cooling, as in  soft-intermediate states, 
the outflow rate increases and the subsonic region has relatively high optical depth (Chakrabarti 1999b).
In some outburst sources, Keplerian matter may rise much faster than the sub-Keplerian flow as in the present case (Paper-I).
Thus, the shock disappears even in HIMS and blobby jets may arise in HIMS as well.

In presence of high Keplerian accretion rates, CENBOL cools down due to high supply of the soft photons from the Keplerian disk. 
Hence it is quenched and we do not see any jet in this state. The results from this considerations are given in 
Fig. 1 (left panel), where the `generic' variation of the ratio of outflow ($\dot{M}_{out}$) and inflow ($\dot{M}_{in}$) 
rate ($R_{\dot{m}}=\frac{\dot{M}_{out}}{\dot{M}_{in}}$) with shock compression ratio ($R$) is shown. Clearly, the ratio ($R_{\dot{m}}$) 
is maximum when the Compression ratio is intermediate as in the hard-intermediate and soft-intermediate states. 
The observed jet in this spectral state is 
dense compact initially, but becomes increasingly blobby as the transition to the soft-intermediate state is approached. 
This is due to the rapid cooling of the jet base, the outflowing matter gets separated since even the 
subsonic flow region becomes suddenly supersonic (Chakrabarti, 1999b; Das \& Chakrabarti, 1999; CM00).

\section{Flux and Spectrum of X-rays from the base of the Jet}

Detailed study of the evolution of the spectral and timing properties for the BHC Swift~J1753.5-0127 during its 2005 outburst 
using the TCAF solution is presented in Paper-I. Depending upon the variation of TCAF model fitted physical flow parameters and 
nature of QPOs (if present), they classified the entire outburst (from  2005 July 2 to 2005 October 19) into two harder (HS and HIMS) 
spectral states. No signatures of softer states (SIMS and SS) were observed. This could be due to the lack of viscosity
that prevented the Keplerian disk to achieve a significant rate close to the black hole.
While fitting spectra with the current version (v0.3) of the TCAF solution, the model normalization (N) is found to vary 
in a very narrow range ($1.41-1.81$), except for a few days when the radio flux was higher. 
This may be because of non-inclusion of the jet mechanism in the current TCAF model 
{\it fits} file. This motivated us to introduce a new method to detect an X-ray jet 
and calculate its contribution from the total X-ray flux.   

We use $2.5-25$~keV RXTE/PCA data to calculate the X-ray flux from the base of the outflow. 
In presence of a jet, the total X-ray flux ($F_{X}$) 
is contributed from the radiation emitted from both the disk and the base of the jet. 
So, during the days with significant X-rays in the outflow, we require higher values of the model normalization
to fit the spectra, since the present version of our TCAF model {\it fits} 
file is only concerned with the emission from the disk and no contribution from the jets is added. 
If the jet is absent, a constant or nearly constant TCAF model normalization is capable 
of fitting the entire outburst (see, Molla et al. 2016, 2017; Chatterjee et al. 2016). In Paper-I, 
TCAF normalization found to be constant at $\sim 1.6$ during the entire 2005 outburst of Swift~J1753.5-0127, except for $5$ observations 
when it assumed higher values ($\ge 2.0$) in the initial period of HIMS (dec.). 
However in HS (dec.) minimum 
normalization of $\sim 1.41$ was required to fit spectral data on 2005 September 17 (MJD=55630.31). We assume that
there was very little X-ray jet or, outflowing matter on that day and the entire X-ray flux is contributed only 
by the accretion disk and CENBOL, i.e., from inflowing matter alone. This is also the theoretical outcome (Chakrabarti, 1999b). 
When we compared the radio data, it was observed that radio flux contributions were also minimum during these days of observations. 
To calculate X-ray flux contribution $F_{inf}$ only from the inflow, we refitted all the spectra by freezing model 
normalization at $1.41$. Then, we take the difference of the resulting spectrum from the total flux to calculate jet X-ray 
flux $F_{ouf}$. In other words, the flux of the jet, relative to MJD=55630.31 can be written as,
$$
F_{ouf} = F_{X} - F_{inf}. \eqno{(1)} 
$$
Here, $F_{X}$ and $F_{inf}$ fluxes (in units of $10^{-9}~ergs~cm^{-2}~s^{-1}$) 
are calculated using `flux 2.5 25.0' command after obtaining the best fitted spectrum in XSPEC. $F_{X}$ is basically 
the TCAF model flux in the energy range of $2.5-25$~keV with free normalization as reported in Paper-I, 
where as $F_{inf}$ is TCAF model flux in the same energy range with constant normalization, N=$1.41$.

\section{Results}

\subsection{Evolution of Jet X-rays}

X-ray fluxes from jets or outflow ($F_{ouf}$) are calculated using Eq. (1).
The variation of the derived jet X-ray flux ($F_{ouf}$) during the 
entire phase of the 2005 outburst of Swift~J1753.5-0127 is shown in Fig. 2(c). To make a comparison, we show the 
variation of $4.8$~GHz VLA radio flux as reported by Soleri et al. (2010) in Fig. 2(d). First radio observation 
was $\sim 5$~days after RXTE/PCA observation, which missed initial two harder spectral states. Note that the 
Radio flux is maximum, during the middle of the HIMS, namely, in the late stage of HIMS in the rising phase and 
early stage of HIMS in the declining phase, precisely as anticipated from the outflow rate behavior in Fig. 1. 
Since the object started to return to the hard state, the outflow rate went down also (Fig. 2c) and thus the radio 
flux also started to go down (Fig. 2d). During the initial $5$~days (MJD=53553.05-53557.24), X-ray flux was completely 
dominated by the inflowing component ($F_{inf}$) and reached its peak on 2005 July 7 (MJD=53557.24), 
which was the day of HS to HIMS transition (Paper-I). Jet X-ray flux ($F_{ouf}$) started
to increase from the transition day and reached its maxima on 2005 July 13 (MJD=53564.91). 
After that the jet X-ray flux starts to decrease; initially the flux reduced rapidly for the next 
$\sim 6$~days and then very slowly or roughly becomes constant until the end of our observation, 
except a weak local peak, observed near on 2005 August 11 (MJD=53593.23).

The TCAF normalization ($N$) also shows a behavior similar to the radio flux of the jet as shown by $F_{ouf}$ plot in Fig. 2c.
It was constant in the first few observations. Then it increased and attained maximum value on the same day when 
$F_{ouf}$ shows peak value on MJD=53564.91. After that, it decreases fast and becomes almost 
constant till the end of our observations, starting from $\sim$ MJD=53570. This additional requirement on $N$
arises from emission of X-rays from the base of the jet, particularly in the subsonic region, which is 
not included in the present version of the TCAF model {\it fits} file. 

The four plots in Fig. 3(a-d) show spectra from four different spectral states (dates marked as online 
red square boxes in Fig. 2e), fitted with free (black solid curve) or frozen 
(online red dashed curve) normalization of the TCAF model. The jet spectrum is also shown 
(online blue dot-dashed curve). It clearly shows that the jet was becoming stronger as the outburst 
progressed and was strongest in HIMS (dec.). Then the contribution from the jet is rapidly 
reduced as the shock receded farther away in the HS (dec.). 

In the strong jet-dominated region (HIMS in the rising and the declining phases), $F_{ouf}$ is observed to be 
in the order of $10^{-9}~ergs~cm^{2}~s^{-1}$, whereas towards the end of the outburst, when 
the jet is weak, it decreases by a factor of a hundred.
We also calculated the contribution of the jet in total X-ray emission. On an average, the flux of X-ray 
jet is $\sim 12.5\%$ of the total X-rays ($F_{X}$). When the jet activity is strong, the contribution rises 
up to $\sim 32\%$ (see, Appendix Table I). The spectrum of X-ray emission from the jet appears to be 
harder than the disk spectrum, which is expected when the base of the jet is optically thin. 
Note also that, the spectral slope of the jet component is different with a turnover property 
at a lower energy than that of the disk as is expected from an expanded system. Though we 
did not plot at lower energy, we expect this region to be downscattered radiation emitted 
from the inflow.

\subsection{Correlation between the Radio and X-ray Jets}

The first radio observation of Swift~J1753.5-0127 was made with MERLIN on 2005 July 3 at $1.7$~GHz (Fender et al. 2005).
WRST and VLA also observed the BHC (Soleri et al. 2010). VLA observed the BHC at $1.4$~GHz, $4.8$~GHz and $8.4$~GHz.
First radio observation was made with VLA on 2005 July 8 (MJD=53558) with radio flux $F_R=2.79$~mJy at $4.8$~GHz.
After that, $F_R$ slightly decreased on $MJD=53561$, before attaining peak on 2005 July 15 (MJD=53566). 
X-ray jets attain its peak roughly two days prior to the radio, i.e., on 2005 July 13 (MJD=53564.91). 
There is $\sim 9$~days gap between 2nd and 3rd radio observations. So, it is hard to find 
exact delay between the X-ray jet and the radio peak fluxes, although there is a gap 
of $\sim 2$~day. Similar to $F_{ouf}$, $F_R$ also showed decreasing nature after 
its peak. $F_R$ decreased rapidly until HIMS (Dec.) to HS (Dec.) transition day (MJD=53589), 
and then decreased slowly and becomes almost constant from $\sim$ MJD=53590. 

It is known from the literature that there exists a correlation between radio and X-ray wave bands from jets.
In Fig. 4(a-d), we draw an $F_R$ versus $F_X$ plot. 
We use the results of the available quasi-simultaneous observations of $4.8$~GHz VLA and 
$2.5-25$~keV RXTE/PCA. 
In an effort to find a relation, we fit the data with $F_R \sim F_{X}^{b}$, where $b$ is a constant. 
In Fig. 4a, we show the variation between jet X-ray ($F_{ouf}$) with radio ($F_R$) from quasi-simultaneous observations. 
We obtained $b \sim 0.59 \pm 0.11$. The relation with the X-ray flux from inflow ($F_{inf}$), shown in Fig. 4b, required 
an index  $b\sim 1.28\pm 0.11$. The relation of soft X-ray ($3-9$ keV) 
and radio (Fig. 4c), which is a standard practice, yields $b\sim1.05 \pm 0.14$. 
When we use $F_R$ and total $F_X$ in the $2.5-25$~keV range, we find $b\sim 1.13 \pm 0.12$ (Fig. 4d).

From these plots, we conclude that the entire X-ray (sum of those from inflow and outflow) is well correlated 
only at lower fluxes be it in $3-9$ keV range (Fig. 4c) or in $2.5-25$ keV range (Fig. 4d). 
However, if we consider outflow X-ray flux ($F_{ouf}$) instead of $F_X$, then the correlation of 
$F_{ouf}$ vs. $F_R$ (Fig. 4a) is found to be weak. However, a good correlation is obtained between $F_R$ 
and X-ray flux from the inflow ($F_{inf}$) at all levels of flux (Fig. 4b). It is possible that the nature 
of the jet deviates from compactness as the intermediate state is approached.
This behavior is compatible with the observed fact that the compact jets are generally 
well correlated with the radio flux, while the blobby jets are not.

Swift~J1753.5-0127 is less luminous in radio as compared to other BHCs (Soleri et al. 2010).
In fact, even during the strong jet observation, the total X-ray flux is not entirely contributed 
by the jets. A large contribution always comes from the accretion disk. 
This may be the reason behind not fitting our result with the standard $b$ ($0.6-0.7$). 
Rushton et al. (2016) also found a similar result. They found the correlation 
index to be $\sim0.99\pm0.12$ in soft (0.6-10 keV) and $\sim0.96\pm0.06$ in hard (15-150 keV) 
X-ray bands using the data of Swift/XRT and Swift/BAT instruments respectively.

\section{Discussions and Concluding Remarks}

In this paper, we use a novel approach to obtain the spectral evolution of the X-rays from the outflow 
component of Swift~J1753.5-0127 during its 2005 outburst by exploiting the fact that the normalization 
of a TCAF fit having X-ray contributions from an inflow remains constant across the states. We use 
$2.5-25$~keV RXTE/PCU2 data of BHC Swift~J1753.5-0127 during its 2005 outburst. 
Much higher normalization values were required to fit spectra on a few days, 
belonging to HIMS (dec.). Assuming the minimum TCAF model normalization, $1.41$ obtained on 2005 September 17 (MJD=55630.31)
to be contributed from the $2.5-25$ keV range flux from accretion flows only, we estimated the outflow contribution 
in rest of the observations. This was done by separating accretion disk spectrum and flux ($F_{inf}$) from the total 
spectrum and flux by refitting all spectra, keeping normalization frozen at $1.41$. X-ray 
flux ($F_{ouf}$) contribution from the outflow was obtained using Eq. 1. Time dependence of X-ray flux and spectrum 
from the outflow thus obtained and the flux variation is appeared to be similar to the observed radio flux data 
(see, Fig. 2d). 

The variations of $F_{inf}$ and $F_{ouf}$ showed that although initially disk flux increased rapidly and attained 
its maximum on 2005 July 7 (MJD=53557.24), the jet flux stays roughly constant. Starting from the time when the 
$F_{inf}$ is maximum, jet flux also starts to increase and attains its maximum on 2005 July 13 (MJD=53564.91) 
when the spectral state changed from hard to hard intermediate. 
In the declining phase, the jet flux decreases and becomes roughly constant in the later phase of the outburst 
and finally became negligible. If we interpret that the radio intensity is directly related to the outflow rate, 
then it should follow the nature of outflow rate (${\dot m} R_{\dot{m}}$, where $R_{\dot{m}}$ variation as in Fig. 1) 
that was predicted by Chakrabarti (1999ab) in the presence of shocks. 
Here, ${\dot m}$ is the sum of the disk and halo component rates that increased
from HS to HIMS (Mondal et al. 2014, 2016; Debnath et al. 2015a,b; Jana et al. 2016; Molla et al. 2017). 

In deriving the properties of the X-rays from the jets, we assumed that the significant variation of the 
TCAF model normalization (N) is entirely due to the variation in jet contribution in X-ray. 
Since the outflow rate is supposed to increase in HIMS, 
it is likely that the X-ray contribution would also go up. We needed $N=2.61$ 
(maximum) on MJD=53564.91 for fitting, when $F_{ouf}$ is observed to be maximum. Correlation between these two is good
until the compactness of the jet is maintained. Higher outflow rates may have caused blobbiness (Chakrabarti, 1999b, 2000) 
and the variation of the outflow contribution with radio was no longer well correlated at higher flux. 
During the radio jet-dominated region, i.e., 
HIMS (dec.), the X-ray jet had a flux of around of $10^{-9}~ergs~cm^{2}~s^{-1}$, 
whereas during the declining phase, the flux drops to $\sim 10^{-11}~ergs~cm^{2}~s^{-1}$, which is about $100$ times 
lower. There are a few examples of X-ray flux measurements of inner jets. For example, Nandi et al. (2005) showed that the 
X-ray flux from the jets for BHC SS~433 is around $10^{-10}~ergs~cm^{2}~s^{-1}$ in $3-25$ keV energy band. 
For 4U~1755-33, X-ray flux from the jet is observed to be around $10^{-16}~ergs~cm^{2}~s^{-1}$ in quiescent state 
(Angelini \& White, 2003). 

In the later part of the 2005 outburst of the BHC Swift J1753.5-0127, radio flux ($F_{R}$) 
was found to be about constant at its lower value ($\sim 0.4$~mJy). 
Toward of the end of our observations, jets may be moderately stronger in radio but weaker in the X-ray band.
Overall, jet X-ray contribution is found to be at $\sim 12.5$\% over the total X-ray. When the jet is strong, i.e., 
in the HIMS, the outflow contribution is about $32$\% of that of the inflow contribution, surprisingly very similar 
to the ratio of the flow rates predicted in HIMS (Chakrabarti, 1999a). 
Our result is consistent with what is observed in other similar compact sources.

In the TCAF solution, the jets are considered to emerge out of CENBOL (Chakrabarti 1999ab), which is the `hot' puffed-up 
region acting as a Compton cloud. The CENBOL acts as the base of the jet. While CENBOL is the post-shock compressed matter 
flowing inward, the matter in the jet is expanding outward and is relatively optically thin. This explains why the spectrum 
from the jet is flatter. As matter expands and interacts with entangled magnetic fields, it emits radio waves, generally 
far away from the black hole. 

Both the X-ray and the radio emissions from outflow depend on the outflow rate. However, X-ray component is strong only 
if the outflow rate is higher as happens when the object goes to HIMS.  Since the shock is weaker, the 
outflow must be radiation driven, rather than thermal pressure driven. The jets could be blobby when the optical 
depth is high and the correlation between the two fluxes breaks down. On the 
other hand, the X-ray emission from the inflow causes $F_{inf}$ to rise also from HS to HIMS. Outflow rate is controlled 
by the shock strength i.e., by the compression ratio $R$ (Fig. 1). Hence, it is expected that a correlation between $F_{inf}$ 
and $F_{R}$ should exist. Since $F_{ouf} << F_{inf}$ this translates to a correlation between total $F_X$ and $F_R$. 
An empirical relation ($F_{R} \propto F_{X}^b$ with $b \sim 0.6-0.7$) was found by Hannikainen et al. 1998; Corbel et al. 2003; 
Gallo et al. 2003), although, some `outliers' were found to have a steeper power-law index ($b \sim 1.4$) (Jonker et al. 2004; 
Coriat et al. 2011). Using quasi-simultaneous observation of VLA at $4.8$~GHz and the $2.5-25$~keV RXTE/PCA TCAF model fitted 
total X-ray flux, we find $b\sim 1.13 \pm 0.12$, for $F_{R}$ and $F_X$ i.e., $F_{R} \propto F_{X}^{1.13 \pm 0.12}$.  
Instead of the $2.5-25$~keV total X-ray flux ($F_X$), using the $3-9$~keV soft X-ray flux, we find a less steep exponent of
$b\sim 1.05\pm 0.14$. Our result is consistent with several other authors, who also have found a steeper exponent for this 
particular BHC with $b\sim 1.0-1.4$ (Soleri et al. 2010, Rushton et al. 2016, Kolehmainen et al. 2016). 
This BHC candidate is less luminous in radio which may be the reason behind getting a steeper index (Soleri et al. 2010).
When $F_{inf}$ and $F_R$ are compared, the index is $\sim 1.28 \pm 0.11$. When $F_{ouf}$ and $F_R$ are compared, $b \sim 0.59 \pm 0.11$. 
The observed points in the high jet-dominated region are not well correlated in the later case ($F_{ouf}$ vs. $F_R$, see Fig. 4a). 
This may be due to the possible blobby nature of the jets in the high flux HIMS (dec.) region of the outburst.

In future, we would like to estimate X-ray jet fluxes for a few other transient BHCs, such as, MAXI~J1836-194, XTE~J1180+480, etc., 
where deviations of the constancy of the TCAF model normalization have been observed (see, Jana et al. 2016; Chatterjee et al. 2016), 
using the same method described in this paper as well as persistent sources such as GRS~1915+105, GX~339-4, V~404~Cyg.

\section*{Acknowledgments}

A.J. and D.D. acknowledge the support from ISRO sponsored RESPOND project fund (ISRO/RES/2/388/2014-15). 
D.D. also acknowledges support from DST sponsored Fast-track Young Scientist project fund (SR/FTP/PS-188/2012).

{}

\clearpage

%%\end{document}
%%%%%%%%%%%%%%%%%%%%%%%%%%%%%%%%%%%%%%%%%%%%%%%%%%%%%%%%%%%%%%%%%%%%%%%%%%%%%%%%%%%%%%

\begin{figure}
\vskip -0.5cm
\centerline{
\includegraphics[scale=0.6,angle=0,width=9.5truecm]{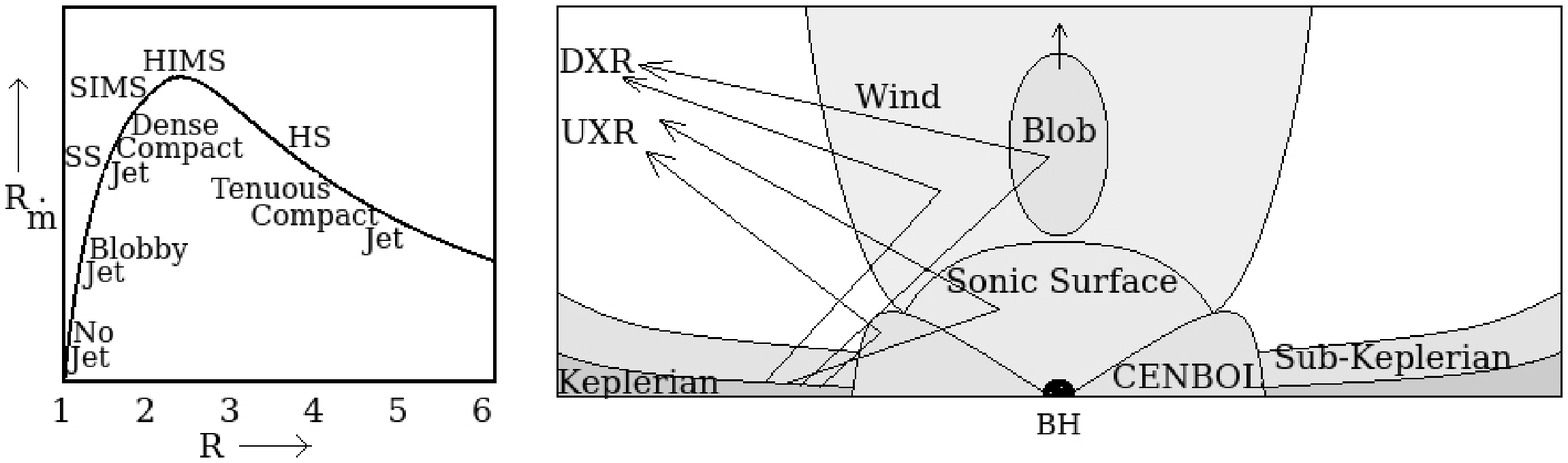}}
\caption{ (Left) Variation of the ratio $R_{\dot m}$ of the outflow and inflow rates as a function of the
compression ratio at the shock $R$. This has a peak at intermediate shock strength. (Right) Examples of
Compton scattering of soft photons off the CENBOL, the subsonic part of the outflow and the blobbs which
may be separated at the intermediate $R$ when the outflow rate is so high that the
optical depth will easily allow significant cooling of the outflow base.}
\label{fig1}
\end{figure}

\begin{figure}
\vskip -0.5cm
\centerline{
\includegraphics[scale=0.6,angle=0,width=8.5truecm]{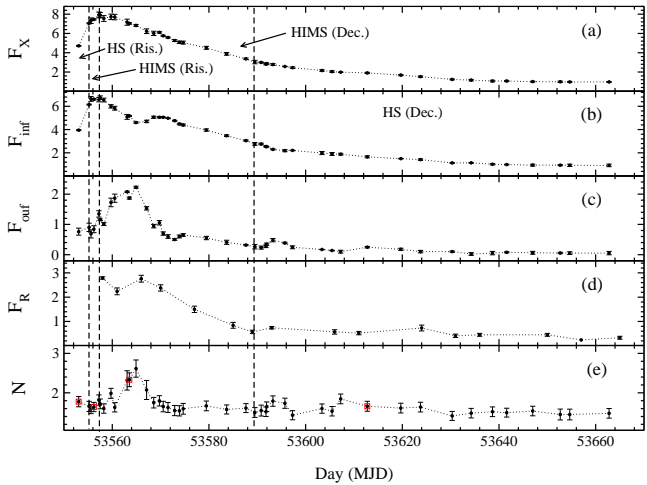}}
\caption{ Variations of (a) total X-ray flux ($F_{X}$), (b) X-ray flux from accretion disk ($F_{inf}$), and (c) from jet ($F_{ouf}$) 
in units of $10^{-9} ergs$ $cm^2$ $s^{-1}$ are shown. In plot (d), variation radio flux ($F_R$ in mJy) in $4.8$~GHz band of VLA 
as reported by Soleri et al. (2010), and in (e), variation of TCAF model fitted normalization (N) are shown with day (MJD).}
\label{fig2}
\end{figure}

%\clearpage

\begin{figure}
\vskip -0.0cm
\centerline{
\includegraphics[scale=0.6,angle=0,width=8.0truecm]{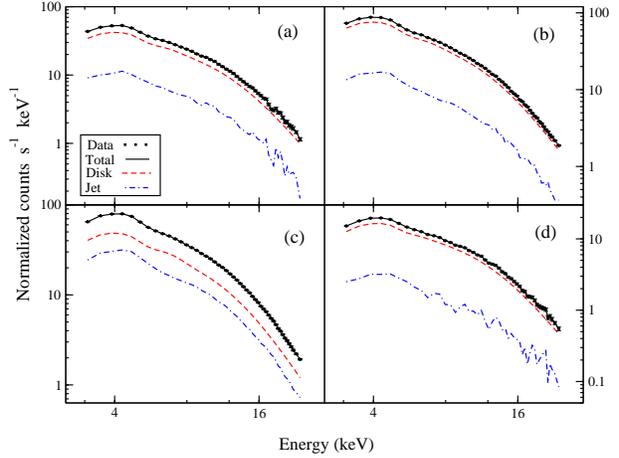}}
\caption{Four spectra selected from four different spectral states, fitted with TCAF model fits file by keeping 
model normalization as free (online blue curve) or frozen at N=1.41 (online red curve). The spectra are from 
observation IDs: (a) 91094-01-01-00, (b) 91094-01-01-03, (c) 91423-01-02-00, and (d) 91423-01-09-00. 
Jet X-ray spectra are shown in online blue dot-dashed curves. 
Note, these spectra are marked as online red square boxes in Fig. 2e.
}
\label{fig3}
\end{figure}

\begin{figure}
\vskip -0.0cm
\centerline{
\includegraphics[scale=0.6,angle=0,width=8.5truecm,height=7cm]{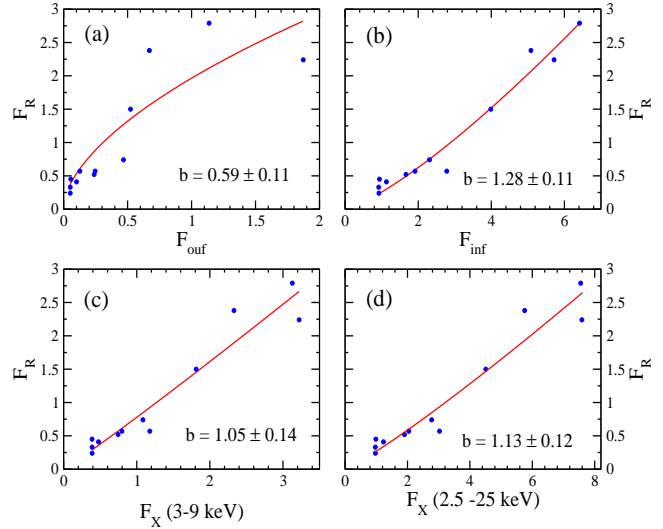}}
\caption{Correlation plot of (a) 2.5-25 keV outflow X-ray ($F_{ouf}$), (b) 2.5-25 keV inflow X-ray ($F_{ouf}$), 
(c) $3-9$~keV total X-ray ($F_X$) and (d) $2.5-25$~keV total X-ray ($F_X$) fluxes with radio (VLA $4.8$~GHz) 
fluxes using quasi-simultaneous observations are shown.}
\label{fig4}
\end{figure}

\clearpage
\begin{table}
\vskip -2.0cm
\addtolength{\tabcolsep}{-4.50pt}
\scriptsize
\centering
\centering{\large \bf Appendix I}
\vskip 0.2cm
\centerline {X-ray flux from Jets}
\vskip 0.2cm
\begin{tabular}{lcccccccc}
\hline
obs& Id&MJD&$F_{X}$&$F_{inf}$&$F_{ouf}$&\% of $F_{ouf}$ &$N$\\
 (1)&  (2)  & (3)  & (4)& (5) & (6) & (7) & (8)\\
\hline

\hline
1&X-01-00& 53553.05 &$4.712^{\pm0.052}$&$3.951^{\pm0.041}$ &$0.761^{\pm0.115}$&$16.14$ &$1.775^{\pm0.130}$ \\
2&X-01-01& 53555.20 &$7.053^{\pm0.049}$&$6.149^{\pm0.035}$ &$0.903^{\pm0.137}$&$12.81$ &$1.655^{\pm0.128}$\\
3&X-01-02& 53555.60 &$7.323^{\pm0.290}$&$6.636^{\pm0.212}$ &$0.687^{\pm0.134}$&$9.379$ &$1.588^{\pm0.117}$\\
4&X-01-03& 53556.19 &$7.451^{\pm0.057}$&$6.610^{\pm0.071}$ &$0.841^{\pm0.116}$&$11.29$ &$1.639^{\pm0.117}$\\
5&Y-01-04& 53557.24 &$7.920^{\pm0.301}$&$6.574^{\pm0.192}$ &$1.345^{\pm0.106}$&$16.98$ &$1.816^{\pm0.129}$\\
6&X-01-04& 53557.50 &$7.898^{\pm0.271}$&$6.742^{\pm0.156}$ &$1.156^{\pm0.027}$&$14.63$ &$1.719^{\pm0.123}$\\
7&Y-01-00& 53558.25 &$7.551^{\pm0.299}$&$6.541^{\pm0.162}$ &$1.010^{\pm0.048}$&$13.37$ &$1.601^{\pm0.118}$\\
8&X-02-01& 53559.73 &$7.721^{\pm0.281}$&$5.995^{\pm0.147}$ &$1.725^{\pm0.141}$&$22.34$ &$1.987^{\pm0.126}$\\
9&X-02-00& 53560.51 &$7.700^{\pm0.283}$&$5.834^{\pm0.167}$ &$1.866^{\pm0.135}$&$24.23$ &$1.632^{\pm0.117}$\\
10&X-02-02& 53561.49 &$7.589^{\pm0.271}$&$5.717^{\pm0.152}$ &$1.872^{\pm0.119}$&$24.67$ &$2.141^{\pm0.119}$\\
11&Y-02-00& 53563.07 &$7.167^{\pm0.315}$&$5.090^{\pm0.209}$ &$2.076^{\pm0.021}$&$28.96$ &$2.315^{\pm0.244}$\\
12&Y-02-05& 53563.52 &$7.040^{\pm0.078}$&$5.172^{\pm0.051}$ &$1.869^{\pm0.033}$&$26.54$ &$2.331^{\pm0.175}$\\
13&Y-02-06& 53564.91 &$6.830^{\pm0.102}$&$4.602^{\pm0.054}$ &$2.228^{\pm0.033}$&$32.62$ &$2.614^{\pm0.218}$\\
14&Y-03-00& 53567.09 &$6.231^{\pm0.242}$&$4.701^{\pm0.102}$ &$1.530^{\pm0.056}$&$24.56$ &$2.072^{\pm0.241}$\\
15&Y-03-02& 53568.57 &$6.006^{\pm0.223}$&$5.064^{\pm0.088}$ &$0.942^{\pm0.058}$&$15.68$ &$1.750^{\pm0.133}$\\
16&Y-03-03& 53569.75 &$6.111^{\pm0.071}$&$5.056^{\pm0.050}$ &$1.054^{\pm0.074}$&$17.25$ &$1.796^{\pm0.130}$\\
17&Y-03-04& 53570.54 &$5.755^{\pm0.061}$&$5.058^{\pm0.028}$ &$0.697^{\pm0.056}$&$12.11$ &$1.652^{\pm0.133}$\\
18&Y-03-05& 53571.52 &$5.579^{\pm0.057}$&$4.980^{\pm0.024}$ &$0.599^{\pm0.061}$&$10.74$ &$1.627^{\pm0.126}$\\
19&Y-03-06& 53572.91 &$5.261^{\pm0.091}$&$4.761^{\pm0.035}$ &$0.500^{\pm0.026}$&$9.510$ &$1.550^{\pm0.129}$\\
20&Y-04-00& 53573.89 &$5.082^{\pm0.122}$&$4.481^{\pm0.064}$ &$0.600^{\pm0.047}$&$11.82$ &$1.542^{\pm0.130}$\\
21&Y-04-01& 53574.67 &$5.049^{\pm0.152}$&$4.394^{\pm0.078}$ &$0.650^{\pm0.044}$&$12.88$ &$1.594^{\pm0.135}$\\
22&Y-04-06& 53579.45 &$4.510^{\pm0.154}$&$3.960^{\pm0.098}$ &$0.550^{\pm0.056}$&$12.20$ &$1.668^{\pm0.123}$\\
23&Y-05-01& 53583.65 &$3.880^{\pm0.140}$&$3.471^{\pm0.071}$ &$0.409^{\pm0.069}$&$10.54$ &$1.577^{\pm0.115}$\\
24&Y-06-00& 53587.64 &$3.367^{\pm0.079}$&$3.050^{\pm0.058}$ &$0.318^{\pm0.021}$&$9.436$ &$1.612^{\pm0.113}$\\
25&Y-06-01& 53589.49 &$3.027^{\pm0.179}$&$2.764^{\pm0.118}$ &$0.263^{\pm0.073}$&$8.696$ &$1.500^{\pm0.125}$\\
26&Y-06-05& 53590.80 &$2.992^{\pm0.045}$&$2.758^{\pm0.071}$ &$0.234^{\pm0.051}$&$7.816$ &$1.550^{\pm0.138}$\\
27&Y-06-06& 53591.79 &$2.824^{\pm0.084}$&$2.544^{\pm0.037}$ &$0.279^{\pm0.045}$&$9.907$ &$1.520^{\pm0.121}$\\
28&Y-06-07& 53591.92 &$2.870^{\pm0.069}$&$2.524^{\pm0.025}$ &$0.345^{\pm0.037}$&$12.05$ &$1.646^{\pm0.122}$\\
29&Y-06-03& 53593.22 &$2.777^{\pm0.097}$&$2.297^{\pm0.041}$ &$0.479^{\pm0.055}$&$17.27$ &$1.790^{\pm0.131}$\\
30&Y-07-00& 53595.71 &$2.572^{\pm0.014}$&$2.185^{\pm0.087}$ &$0.387^{\pm0.023}$&$15.03$ &$1.735^{\pm0.128}$\\
31&Y-07-01& 53597.29 &$2.448^{\pm0.075}$&$2.205^{\pm0.024}$ &$0.244^{\pm0.048}$&$9.951$ &$1.432^{\pm0.115}$\\
32&Y-08-01& 53603.38 &$2.155^{\pm0.087}$&$1.984^{\pm0.132}$ &$0.171^{\pm0.006}$&$7.919$ &$1.600^{\pm0.117}$\\
33&Y-08-02& 53605.46 &$2.042^{\pm0.088}$&$1.905^{\pm0.125}$ &$0.137^{\pm0.007}$&$6.713$ &$1.533^{\pm0.119}$\\
34&Y-08-03& 53607.25 &$1.982^{\pm0.022}$&$1.882^{\pm0.077}$ &$0.099^{\pm0.048}$&$5.009$ &$1.846^{\pm0.123}$\\
35&Y-09-00& 53612.75 &$1.903^{\pm0.064}$&$1.655^{\pm0.087}$ &$0.247^{\pm0.025}$&$13.01$ &$1.658^{\pm0.126}$\\
36&Y-10-00& 53619.69 &$1.678^{\pm0.069}$&$1.501^{\pm0.021}$ &$0.177^{\pm0.036}$&$10.57$ &$1.612^{\pm0.123}$\\
37&Y-11-00& 53623.76 &$1.518^{\pm0.071}$&$1.417^{\pm0.065}$ &$0.101^{\pm0.041}$&$6.649$ &$1.638^{\pm0.122}$\\
38&Y-12-00& 53630.31 &$1.229^{\pm0.049}$&$1.130^{\pm0.042}$ &$0.100^{\pm0.019}$&$8.188$ &$1.414^{\pm0.114}$\\
39&Y-12-01& 53634.24 &$1.156^{\pm0.079}$&$1.138^{\pm0.031}$ &$0.022^{\pm0.045}$&$1.901$ &$1.484^{\pm0.121}$\\
40&Y-13-00& 53638.64 &$1.064^{\pm0.081}$&$1.016^{\pm0.056}$ &$0.052^{\pm0.056}$&$4.876$ &$1.519^{\pm0.123}$\\
41&Y-13-01& 53641.59 &$1.066^{\pm0.063}$&$0.987^{\pm0.027}$ &$0.078^{\pm0.017}$&$7.361$ &$1.496^{\pm0.111}$\\
42&Y-14-00& 53646.97 &$1.005^{\pm0.042}$&$0.942^{\pm0.083}$ &$0.063^{\pm0.047}$&$6.258$ &$1.538^{\pm0.121}$\\
43&Y-15-00& 53652.66 &$0.985^{\pm0.071}$&$0.941^{\pm0.052}$ &$0.055^{\pm0.023}$&$5.640$ &$1.450^{\pm0.137}$\\
44&Y-15-01& 53654.62 &$0.973^{\pm0.054}$&$0.924^{\pm0.099}$ &$0.051^{\pm0.045}$&$5.301$ &$1.450^{\pm0.136}$\\
45&Y-16-01& 53662.76 &$0.969^{\pm0.052}$&$0.921^{\pm0.108}$ &$0.052^{\pm0.056}$&$5.332$ &$1.480^{\pm0.122}$\\
 
\hline
\end{tabular}
\noindent{
\leftline {X=91094-01 and Y=91423-01 are the prefixes of observation Ids.} 
\leftline { Total X-ray flux,accretion disk X-ray flux and Jet X-ray flux are in the units of $10^{-9}$ $ergs~cm^{-2}~s^{-1}$.
$F_X$, $F_{inf}$ and $F_{ouf}$ are calculated in 2.5-25 keV energy range.} 
\leftline {Note: average values of 90\% confidence $\pm$ values obtained using `err' task in XSPEC, 
are placed as superscripts of fitted parameter values.}
}
\end{table}

 %%%%%%%%%%%%%%%%%

\begin{thebibliography}{}

\bibitem[Angelini \& White 2003]{AW03} Angelini, L., \& White, N. E. 2003, ApJ, 586, L71
\bibitem[Bhattacharjee et al.(2017)]{B17} Bhattacharjee, A., Banerjee, I., Banerjee, A. et al., 2017, MNRAS, 466, 1372
\bibitem[Cadolle Bel et al. 2007]{Cadolle Bel 2007} Cadolle Bel, M., Rib\'{o}, M., Rodriguez, J. et al. 2007, ApJ, 659, 549
\bibitem[Chakrabarti \& D'Silva(1994)]{CD94} Chakrabarti, S. K., \& D'Silva, S. 1994, ApJ, 424, 138
\bibitem[Chakrabarti \& Titarchuk(1995)]{CT95} Chakrabarti, S. K., \& Titarchuk, L.G. 1995, ApJ, 455, 623 (CT95)
\bibitem[Chakrabarti(1999a)]{C99a} Chakrabarti, S. K. 1999a, A\&A, 351, 185
\bibitem[Chakrabarti(1999b)]{C99b} Chakrabarti, S. K. 1999b, Ind. J. Phys., 73B, 6, 931
\bibitem[Chakrabarti \& Manickam (2000)]{CM00} Chakrabarti, S, K, \& Manickam, S. G. 2000, ApJ, 531, L41
\bibitem[Chakrabarti 2000]{C00} Chakrabarti, S. K. 2000, CQGra, 17, 2427
\bibitem[Chakrabarti, \& Nandi,(2000)]{CN00} Chakrabarti, S. K., \& Nandi, A., 2000, Ind. J. Phys. 75(B), 1 (arxiv:12526)
\bibitem[Chakrabarti(2001)]{C01} Chakrabarti, S. K. 2001, AIPC, 558, 831
\bibitem[Chakrabarti(2002)]{CGW02} Chakrabarti, S. k., Goldoni, P., Wiita, P. J. et al. 2002, ApJ, 576, L45
\bibitem[Chakrabarti(2013)]{c13} Chakrabarti, S. K., 2013, Astron.  Soc. of India Conf. Ser., 8, 1
\bibitem[Chatterjee et al.(2016)]{C16} Chatterjee, D., Debnath, D., Chakrabarti, S.K, et al. 2016, ApJ, 827, 88
\bibitem[Corbel2002a]{Corbel2002a} Corbel, S., Fender, R., \& Tzioumis, A. 2002a, IAU Circ. 7795, 2
\bibitem[Corbel200b]{Corbel2002b} Corbel, S., Fender, R., \& Tzioumis, A. 2002b, Science, 298, 196
\bibitem[Corbel et al.(2003)]{Corbel03} Corbel, S., Nowak, M. A., Fender, R. P., Tzioumis, A. K., \& Markoff, S. 2003, A\&A, 400, 1007
\bibitem[Corbel et al.(2004)]{Corbel04} Corbel, S., Fender, R. P., Tomsick, J. A., Tzioumis, A. K., \& Tin-gay, S. 2004, ApJ, 617, 1272
\bibitem[Corbel et al.(2005)]{Corbel05} Corbel, S., Kaaret, P., Fender, R. P. et al. 2005, ApJ, 632, 504
\bibitem[Coriat et al.(2011)]{Coriat11} Coriat, M., Corbel, S., Prat, L., et al., 2011, MNRAS, 414, 677
\bibitem[Das \& Chakrabarti(1999)]{DC99} Das, T. K. \& Chakrabarti, S. K. 1999, CQGra, 16, 3879
\bibitem[Debnath et al.(2008)]{DD08} Debnath, D., Chakrabarti, S.K., \& Nandi, A. et al. 2008, BASI, 36, 151
\bibitem[Debnath et al.(2010)]{DD10} Debnath, D., Chakrabarti, S. K., \& Nandi, A. 2010, A\&A, 520, A98
\bibitem[Debnath et al.(2013)]{DD13} Debnath, D., Chakrabarti, S. K., \& Nandi, A. 2013, AdSpR, 52, 2143
\bibitem[Debnath, Chakrabarti \& Mondal(2014)]{DCM14} Debnath, D., Chakrabarti, S. K., \& Mondal, S. 2014, MNRAS, 440, L121
\bibitem[Debnath, Mondal \& Chakrabarti(2015a)]{DMC15} Debnath, D., Mondal, S., \& Chakrabarti, S. K. 2015a, MNRAS, 447, 1984
\bibitem[Debnath et al.(2015b)]{DMCM15} Debnath, D., Molla, A. A., Chakrabarti, S. K. \& Mondal, S. 2015b, ApJ, 803, 59
\bibitem[Debnath et al.(2017)]{DD17} Debnath, D., Jana, A., Chakrabarti, S. K. et al. 2017, ApJ (submitted) (arXiv:1703.05479) (Paper-I)
\bibitem{Dhawan01}Dhawan, V., Mirabel, I. F., \& Rodr\'{i}guez, L. F. 2000, ApJ, 543, 373
\bibitem[D'Silva \& Chakrabarti(1994)]{DC94} D'Silva, S., Chakrabarti, S. K. 1994, ApJ, 424, 149
\bibitem[Eikenberry et al. 1998]{Eikenberry98} Eikenberry, S. S., Matthews, K., \& Morgan, H. et al. 1998, ApJ, 494, L61
\bibitem[Fender et al.(2004)]{Fender04} Fender, R. P., Belloni, T. M., Gallo, E. 2004, MNRAS, 355, 1105
\bibitem[Fender et al.(2005)]{Fender05} Fender, R. P., Garrington, S., Muxlow, T. 2005, ATel, 558, 1
\bibitem[Garain et al. (2013)]{Garain13} Garain, S. K., Ghosh, H., \& Chakrabarti, S. K. 2013, ASInC, 8, 11
\bibitem[Gallo et al.(2003)]{Gallo03}Gallo, E., Fender, R. P., \& Pooley, G. G. 2003, MNRAS, 344, 60
\bibitem[Hannikainen et al.(1998)]{Hannikainen98} Hannikainen, D. C., Hunstead, R. W., Campbell-Wilson, D., \& Sood, R. K. 1998, A\&A, 337, 460
\bibitem[Hjellming95]{HR95}Hjellming, R. M., \& Rupen, M. P. 1995, Nature, 375, 464
\bibitem[Jana,Debnath \& Chakrabarti(2016)]{Jana16} Jana, A., Debnath, D., Chakrabarti, S. K. et al. 2016, ApJ, 803, 107
\bibitem[Jonker et al.(2004)]{Jonker04} Jonker, P. G., Gallo, E., Dhawan, V. et al. 2004, MNRAS, 351, 1359
\bibitem[Jonker et al.(2012)]{Jonker12} Jonker, P. G., Miller-Jones, J. C. A., Homan, J. et al. 2012, MNRAS, 423, 3308
\bibitem[Jourdain et al. (2012)]{Jourdain12} Jourdain, E., Roques, J. P., Chauvin, M., \& Clark, D. J., 2012, ApJ, 761, 27
\bibitem[Kaaret,2006]{Kaaret06} Kaaret, P., Corbel, S., \& Tomsick, J. A. et al. 2006, ApJ, 641, 410
\bibitem[Kolehmainen et al.(2016)]{Kolehmainen16} Kolehmainen, M., Fender, R., Jonker, P.G., et al. 2016, AN, 337, 485
\bibitem[Lasso-Cabrera 2006]{LC06} Lasso-Cabrera, N. M., Eikenberry, S. S. 2013, ApJ, 775, 82
\bibitem[Laurent et al. (2011)]{L11} Laurent, P., Rodriguez, J., \& Wilms, J. er at. 2011, Sci, 332, L438
\bibitem[Loh et al. (2016)]{Loh16} Loh, A., Corbel, S., \& Dubus, G., et al. 2016, MNRAS, 462, L111
\bibitem{Mirabel92}Mirabel, I. F., Rodr\'{i}guez, L. F., Cordier, B., Paul, J., \& Lebrun, F. 1992, Nature, 358, 215
\bibitem[Mirabel94]{Mirabel94} Mirabel, I. F., Rodr\'{i}guez, L. F. 1994, Nature, 371, 46
\bibitem[Molla et al. 2017]{Molla17} Molla, A. A., Debnath, D., \& Chakrabarti, S. K. et al. 2017, ApJ, 834, 88
\bibitem[Molla et al.(2016)]{Molla16} Molla, A. A., Debnath, D., \& Chakrabarti, S. K. et al. 2016b, MNRAS, 460. 3163 
\bibitem[Mondal, Debnath \& Chakrabarti(2014a)]{Mondal14a} Mondal, S., Debnath, D., \& Chakrabarti, S. K. 2014, ApJ, 786, 4
\bibitem[Mondal, Chakrabarti \& Debnath(2014b)]{Mondal14b} Mondal, S., Chakrabarti, S.K., \& Debnath, D. 2014b, Ap$\&$SS, 353, 223
\bibitem[Mondal, Chakrabarti \& Debnath(2015)]{Mondal15} Mondal, S., Chakrabarti, S.K., \& Debnath, D. 2015, ApJ, 798, 57
\bibitem[Mondal, Chakrabarti \& Debnath(2016)]{Mondal16} Mondal, S., Chakrabarti, S. K., \& Debnath, D. 2016, Ap$\&$SS, 361, 309
\bibitem[Nandi (2001)]{Nandi2001} Nandi, A., Chakrabarti, S. K., Vadawale, S. V., \& Rao, A. R. 2001, A\&A, 380, 245
\bibitem[Nandi et al.2005]{Nandi05} Nandi, A., Chakrabarti, S. K. \& Belloni, T. et al. 2005, MNRAS, 359, 629
\bibitem[Nandi et al.(2012)]{Nandi12} Nandi, A., Debnath, D., Mandal, S., \& Chakrabarti, S.K., 2012, A\&A, 542, A56
\bibitem[Neustroev et al. 2014]{Neustroev14} Neustroev, V. V., Veledina, A, Poutanen, J. et al. 2014, MNRAS, 445, 2424
\bibitem[Palmer et al. (2005)]{Palmer2005} Palmer, D. M., Barthelmey, S. D. \& Cummings, J. R. et al. 2005, ATel, 546, 1
\bibitem[Ratti et al.(2012)]{Ratti12} Ratti, E. M., Jonker, P. G., Miller-Jones, J. C. A. et al. 2012, MNRAS, 423, 2656
\bibitem{RM92}Rodr\'{i}guez, L. F., Mirabel, I. F., \& Marti, J. 1992, ApJ, 401, L15
\bibitem[Rushton et al. (2016)]{Rushton16} Rushton, A. P.; Shaw, A. W.; Fender, R. P. et al. 2016, MNRAS, 463, 628
\bibitem[Shakura \& Sunyaev(1973)]{SS73} Shakura, N. I., \& Sunyaev, R. A. 1973, A\&A, 24, 337
\bibitem[Shaw et al.(2016)]{Shaw16} Shaw, A. W., Charles, P. A., Casares, J., Hern\'{a}ndez Santisteban, J. V. 2016, MNRAS, 463, 1314
\bibitem[Soleri et al.(2010)]{Soleri10} Soleri, P., Fender, R. P., Tudose, V. et al. 2010, MNRAS, 406, 1471
\bibitem[Sunyaev \& Titarchuk(1980)]{ST80} Sunyaev, R.A., \& Titarchuk, L.G. 1980, ApJ, 86, 121
\bibitem{Stirling01}Stirling, A. M., Spencer, R. E., de la Force, et al. 2001, MNRAS, 327, 1273
\bibitem[Tomsick et al.(2003)]{Tomsick03} Tomsick, J. A., Corbel, S. \& Fender., R. et al. 2003, ApJ, 582, 983
\bibitem{Tingay95}Tingay, S. J., et al. 1995, Nature, 374, 141
\bibitem[Zurita et al.(2007)]{Zurita07} Zurita, C., Torres, M.A.P., Durant, M., et al. 2007, ATel, 1130, 1

\end{thebibliography}
\end{document}